\documentclass[pra,showpacs,floatfix,superscriptaddress,twocolumn]{revtex4}
\usepackage[T1]{fontenc}
\usepackage{amsmath,amsfonts,amssymb,float,graphicx,epsfig,color,times,bbm}
\usepackage{ifsym}
\newcommand{\id}{{\sf 1 \hspace{-0.3ex} \rule{0.1ex}{1.52ex}\rule[-.01ex]{0.3ex}{0.1ex}}}
\begin{document}
\title{Probing the quantum phase transition in the Dicke model through mechanical vibrations}
\author{J. P. Santos}
\affiliation{Departamento de Física, Universidade Estadual de Ponta Grossa - Campus Uvaranas, 84030-900 Ponta Grossa, Paraná, Brazil}
\author{K. Furuya}
\affiliation{Instituto de Física ``Gleb Wataghin'',Universidade Estadual de Campinas - UNICAMP,
Rua Sergio Buarque de Holanda 777, 13083-859, Campinas, Brazil} 
\author{F. L. Semi\~ao}
\affiliation{Departamento de Física, Universidade Estadual de Ponta Grossa - Campus Uvaranas, 84030-900 Ponta Grossa, Paraná, Brazil}
\begin{abstract}
This paper is concerned with quantum dynamics of a system coupled to a critical reservoir. In this context, we employ the Dicke model which is known to exhibit a super radiant quantum phase transition (QPT) and we allow one of the mirrors to move under a linear restoring force.  The electromagnetic field couples to the movable mirror though radiation pressure just like in typical optomechanical setups. We show that, in the  thermodynamical limit, the super-radiant phase induces a classical driving force on the mirror without causing decoherence. 
\end{abstract}
\maketitle
\section{Introduction}\label{intro}
Quantum phase transitions (QPTs) are qualitative modifications that occur in interacting quantum many body systems at zero temperature. They are driven by quantum fluctuations that are induced by a change either of coupling constants or external parameters of the system \cite{sachdev99,qptreview97}. The criticality in the phenomena of phase transition is in general governed by a diverging correlation length and, specially in the case of QPT, it has been conjectured that a quantum correlation with no classical counterpart - the entanglement - must play a crucial role \cite{QPTentang02}.  The methods used to study phase transitions in statistical mechanics required the introduction of new concepts from quantum information theory to characterize and classify the critical ground states at the QPT \cite{amico08}. Seminal articles on spin systems undergoing QPT \cite{QPTentang02,qptSpin} have motivated an enormous amount of work aiming at properly defining the role of entanglement in QFTs. For example, the occurrence of some QPT are indicated by non-analyticity of entanglement measures  near the quantum critical point \cite{wu04,amico08,deoliveira06}. However, there is no one-to-one correspondence between quantum phase transitions and the non-analyticity of entanglement measures when approaching the quantum critical point \cite{yang}.

Among the most studied systems presenting collective quantum phenomena in atomic physics and quantum optics, the Dicke model (DM) stands out \cite{dicke,HB}. It comprises a system of $N$ two-level atoms interacting with a single quantized electromagnetic mode, and it undergoes the QFT known as super-radiant phase transition \cite{andreev93,benedict96,rehler71}. The finite temperature phase transition in the thermodynamic limit of large-$N$ and weak coupling of DM was in the seventies \cite{hepp73,wang73}. More recently, the role of this QFT on quantum chaos signatures has been analyzed  \cite{lewenkopf91,aguiar91,aguiar92,emary03,hou04}, the exact expression for the scaled concurrence obtained \cite{lambert04}, and the scaling of relevant physical quantities determined \cite{vidal06,liu09}. Universality of these results has been extended to the case of inhomogeneous coupling \cite{varcoup}. 
On the practical side, some alternative physical implementations of effective DM have been proposed in different settings such as atom-optical systems with multilevel atoms and cavity-mediated
Raman transitions \cite{car_par}, or even solid-state devices involving superconducting qubits \cite{solidstDM}. 

Another recent issue related to QPT is the question of how a system coupled to a critical bath will behave as the bath undergoes a QPT. Topics such as decoherence, the decay of Loshmidt echo, and entanglement have been studied in systems coupled to a surrounding environment that presents a QPT \cite{sun,sun09,zurek09,zhang09,Li,Yi}. In this article, we address this issue of quantum dynamics of a system coupled to critical reservoir. We consider the Dicke model as a structured bath to which a further oscillator is coupled. For this choice, we were inspired by typical optomechanical systems where the electromagnetic field couples to the mechanical vibrational motion of a mirror subjected to a linear restoring force \cite{optom,nature01,driven}, and we attempt to answer the question: \emph{What happens to the moving mirror when the field and atoms enclosed in the the cavity undergo the super radiant phase transition?} We show that this QFT induces an effective classical \emph{non fluctuating} driving force on the mirror. This result differs from the common case of a bath of harmonic oscillators, for example, where the presence of the reservoir causes a \emph{fluctuating} driving force responsible for the decoherence of the system. 

This article is organized as follows. In Sec. \ref{model}, we describe the optomechanical system used in this work. In Sec. \ref{results}, we discuss the thermodynamical limit of this system, and  present the response of the moving mirror to the super radiant quantum phase transition. In Sec. \ref{final}, we summarize our results. Semiclassical calculations are  shown in the Appendix. 
%
\section{Model Setup}\label{model}
In the Dicke model, an ensemble of $N$ two-level atoms {\bf is} coupled to a single-mode quantized electromagnetic field in a cavity. Such a model contains a QFT from the normal phase to the super-radiant phase through variation of the strength of the coupling constants between the field and the atoms  \cite{andreev93,benedict96,rehler71}.  The model is described by the Dicke Hamiltonian which is given by \cite{dicke,HB} ($\hbar=1$) 
\begin{eqnarray}\label{D}
H_D=\omega a^\dag a+\omega_0J_z+\frac{\lambda}{\sqrt{N}}(a^\dag+a)(J_++J_-),
\end{eqnarray}
where 
\begin{eqnarray}\label{ome}
\omega=\frac{n\pi C}{L}\,\,\,\,\,\,\,(n=1,2,3,\ldots)
\end{eqnarray}  
is the angular frequency of the electromagnetic field mode defined by fixing $n$ (with creation and  annihilation operators $a^\dag$ and $a$, respectively), $C$ the speed of light, $L$ distance between the mirrors, $\omega_0$ is the level-splitting of the two-level atoms, $\{J_+, J_-, J_z\}$ the collective atomic operators satisfying the usual commutation relations of angular momentum, and
\begin{eqnarray}\label{lamb}
\lambda=\wp\mathcal{E}\sin(kx_0),
\end{eqnarray} 
is the coupling constant between the atoms and the cavity field, with $\wp$ the atomic electric dipole matrix element,  $\mathcal{E}=\sqrt{\omega/\epsilon_0L}$ the r.m.s. vacuum electric field amplitude, $\epsilon_0$ the vacuum electric permittivity constant,  and $x_0$ the position of the atoms in the cavity axis.  

Notice that we have assumed identical coupling constants $\lambda$ (same position $x_0$) for all atoms.  This means that we are considering an atomic gas confined in a region of small extent as compared with the cavity size $L$. This assumption is standard in the treatment of the Dicke model \cite{emary03,lambert04,sun}. For a more general setup where the atoms spread in a considerable region of the cavity (gas of large extent), it turns out that the critical behavior is essentially
the same as that in the ordinary model as long as the calculations are done with the average coupling constant  \cite{varcoup}. It is worthwhile to mention that the generalization for a traveling quantized field presents an even richer phase structure when a gas of large extent is considered \cite{cpsun01}, and that the inclusion of atomic motion leads
to modified normal and superradiant phases \cite{larson}. 

\begin{figure}[ht]
 \centering\includegraphics[width=0.8\columnwidth]{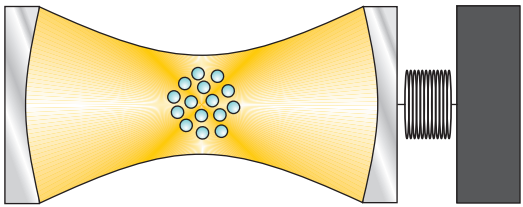}
 \caption{(Color online) A collection of $N$ two-level atoms interact with a single-mode quantized cavity field.  The cavity has one of its mirrors fixed and the other is allowed to move under the action of a linear restoring force which causes vibration of the moving mirror.}
 \label{fig01}
\end{figure}

We now consider that one of the mirrors is allowed to move under the action of a linear restoring force, as depicted in Fig. \ref{fig01}. This can be achieved, for example, by mounting the mirror on a cantilever. The strong coupling regime between a mode of the cavity electromagnetic field and the vibrational motion of the mirror has already been experimentally realized \cite{nature01}. 
Let us denote the angular frequency of vibrational motion by $\omega_m$, and the associated creation and annihilation operators by $c^\dag$ and $c$, respectively. This motion induces the expected optomechanical radiation-pressure interaction which couples the cavity photon number $a^\dag a$ to the position
$X_m=(c^\dag+c)/\sqrt{(2m\omega_m)}$ of the mechanical oscillator, where $m$ the effective mirror mass. However, the mirror vibration also induces a three-body interaction as previously noted for one-atom case \cite{cpsun02}. The physical reason behind the appearance of this coupling is that once the mirror moves, the spatial profile of the cavity field is slightly modified causing changes on the strength of the atom-field coupling $\lambda$. Let us now derive such Hamiltonian. Due to the mirror motion, the cavity changes it length from $L$ to $L+X_m$. Considering  the oscillations very small as
 compared to $L$, Hamiltonian (\ref{D}) becomes 
\begin{eqnarray}
H&=&\frac{n\pi c}{L}\left(1-\frac{X_m}{L}\right) a^\dag a+\omega_m c^\dag c+\omega_0J_z\nonumber\\ &&+\frac{\wp\mathcal{E}}{\sqrt{N}}[\sin (k x_{0})- X_m\delta](a^\dag+a)(J_++J_-),\nonumber\\
\end{eqnarray} 
where $k=n\pi/L$ and $\delta=[\sin(kx_0)+kx_0\cos(kx_0)]/L$, and we also included the free term $\omega_m c^\dag c$.
We notice that besides the radiation-pressure interaction part, a three-body interaction term also appears. Let us rewrite the above Hamiltonian as
\begin{eqnarray}
H&=&\omega a^\dag a+\omega_m c^\dag c+\omega_0J_z+\frac{\lambda}{\sqrt{N}}(a^\dag+a)(J_++J_-)\nonumber\\ &&-ga^\dag a(c^\dag+c)-\frac{\eta}{\sqrt{N}} (c^\dag+c)(a^\dag+a)(J_++J_-),\nonumber\\
\end{eqnarray}
where $g=\omega/[L\sqrt{(2m\omega_m)}]$ and $\eta=\wp\mathcal{E}\delta/\sqrt{(2m\omega_m)}$. Depending on the specific choices of the system parameters, one or another term may dominate. We choose to work with the two-body radiation-pressure term, since it has already been experimentally realized \cite{nature01}. To compare the relative importance of these terms, we need to consider $\eta/g=\frac{\mathcal{E}\wp}{\omega L}[\sin(kx_0)+kx_0\cos(kx_0)]$. For example, for the fundamental mode $n=1$, we can see that $\eta/g=0$ for $kx_0\approx 2.02876$. This means positioning the atoms around $2L/3$. With this choice, the system Hamiltonian reduces to
\begin{eqnarray}
H&=&\omega a^\dag a+\omega_m c^\dag c+\omega_0J_z+\frac{\lambda}{\sqrt{N}}(a^\dag+a)(J_++J_-)\nonumber\\ &&-\frac{g_0}{N}a^\dag a(c^\dag+c),\nonumber\\
\label{hamiltonian}
\end{eqnarray}
where we define $g_0=gN$. Now the dependence of the radiation pressure coupling on $N$,  arising from the density of atoms $\rho=N/L$, appears explicitly in the Hamiltonian.
\section{Effective driving force on the mirror}\label{results}
We are interested in the thermodynamical limit of this Hamiltonian, where the QFT effectively takes place. The mathematical procedure to reach this limit in each phase of the Dicke model is  thoroughly explained in \cite{emary03}, therefore, we shall present here only the points needed to understand our model which includes an extra oscillator.

First, we perform the Holstein-Primakoff transformation defined as
\begin{eqnarray}
J_+&=&b^\dag\sqrt{N-b^\dag b}\\
J_z&=&b^\dag b-\frac{N}{2},
\end{eqnarray}
where the set of atoms is described by  bosonic operators $b^{\dagger},b$,
 satisfying 
$[b,b^\dag]=\id$. After performing this transformation, the system Hamiltonian becomes (up to constant terms)
\begin{eqnarray}\label{ho}
H&=&\omega a^\dag a+\omega_0b^\dag b+\omega_m c^\dag c-\frac{g_0}{N}a^\dag a(c^\dag+c)\nonumber\\ &&+\lambda(a^\dag+a)\left(b^\dag\sqrt{1-\frac{b^\dag b}{N}}+\sqrt{1-\frac{b^\dag b}{N}}b\right).
\end{eqnarray}
 
Now the system Hamiltonian is in a suitable form for the thermodynamic limit to be applied. The Dicke Hamiltonian presents two phases, and the physical features of these phases motivates the mathematical treatment of the thermodynamical limit. In the normal phase $\lambda<\lambda_c$, where $\lambda_c=\sqrt{\omega\omega_0}/2$, the simple application of the limit in the Hamiltonian (\ref{ho}) leads directly to the correct effective Hamiltonian of the \emph{normal phase}
\begin{eqnarray}\label{np}
H_{\rm{np}}=H_{\rm{np}}^{\rm{Dicke}}+\omega_m c^\dag c,
\label{hnp}
\end{eqnarray}
where 
\begin{eqnarray}
H_{\rm{np}}^{\rm{Dicke}}=\omega a^\dag a+\omega_0b^\dag b+\lambda(a^\dag+a)(b^\dag+b).
\end{eqnarray}
We can see that in the normal phase, the moving mirror effectively \emph{decouples} from the Dicke system and evolves as a \emph{free harmonic oscillator}. We interpret this as a consequence from the fact that the Dicke system is only microscopically excited in the normal phase. 

On the other hand, for $\lambda>\lambda_c$ we must incorporate the fact that both the field and the
atomic gas acquire macroscopic occupations. This motivates the displacement of the bosonic modes as $a^\dag\rightarrow a^\dag+\sqrt{\alpha}$ and $b^\dag\rightarrow b^\dag-\sqrt{\beta}$ \cite{emary03}. If the displacement constants are chosen as
\begin{eqnarray}
\sqrt{\alpha} &=&\frac{\lambda}{\omega}\sqrt{N(1-\mu^2)}\\
\sqrt{\beta} &=& \sqrt{\frac{N}{2}(1-\mu)},
\end{eqnarray}
where $\mu=\omega\omega_0/4\lambda^2=(\lambda_c/\lambda)^2$, and we set the terms with overall
powers of $N$ in the denominator to zero after expanding the  square roots in Hamiltonian (\ref{ho}) considering large $N$, we obtain (up to constant terms)
\begin{eqnarray}\label{srp}
H_{\rm{srp}}&=&H_{\rm{srp}}^{\rm{Dicke}}+\omega_m c^\dag c-\frac{g_0\lambda^2}{\omega^2}(1-\mu^2)(c^\dag+c),
\label{hsrp}
\end{eqnarray}
where 
\begin{eqnarray}
H_{\rm{srp}}^{\rm{Dicke}}&=&\omega a^\dag a+\frac{\omega_0}{2\mu}(1+\mu)b^\dag b+\nonumber\\
&&+\frac{\omega_0(1-\mu)(3+\mu)}{8\mu(1+\mu)}(b^\dag+b)^2\nonumber\\
&&+\lambda\mu\sqrt{\frac{2}{1+\mu}}(a^\dag+a)(b^\dag+b).
\end{eqnarray}
Therefore, in the \emph{super-radiant phase} the moving mirror also effectively decouples from the Dicke system, but it suffers the consequences of the field and atoms being macroscopically excited. This  is translated as an effective classical driving force acting on the mirror. 

This is a very interesting point because an ordinary thermal reservoir would never drive coherent oscillations, even at zero temperature. There is a special combination of facts leading to this particular effect in this optomechanical setup as we are now going to explain. It is well known that the coupling to a quantum environment results in a \emph{fluctuating} force acting on the system and reflecting the characteristics of the reservoir \cite{dissip}. It is precisely the nature of the force fluctuations that cause decoherence and damping. On the other hand, if the number of elements in the reservoir is small, energy can be transfered from the system to the reservoir and then fed back to the system (Poincar\'e recurrence) in a period relevant to the natural time scales of the problem. In cases where the system is coupled to a huge reservoir, the Poincar\'e recurrence time will tend to infinity. When thinking in terms of a reservoir-induced fluctuating force on the system, the former case corresponds to a practically not-fluctuating force. In the optomechanical system treated here, the mirror is coupled solely to a \emph{single member} of the elements composing the critical reservoir, namely the electromagnetic field. Therefore, Poincar\'e recurrences will take place, explaining the energy oscillation of the mirror system, i.e., the fact that $c^\dag c$ is no longer a constant of movement. This still does not explain why the mirror does not suffer minimal decoherence, something expected when the reservoir is traced out, even if it contains only a single element. The point is that the atoms work as a reservoir for the field causing it to behave classically, i.e., $a\rightarrow\sqrt{\alpha}$ such that
\begin{eqnarray}
\frac{g_0}{N}a^\dag a(c^\dag+c)\rightarrow \frac{g_0}{N}\alpha( c^\dag+ c).
\end{eqnarray}  
This is consistent with the fact that the macroscopic excitation in the field is given by $\langle a^\dag a\rangle/J=\alpha/J$ (with $2J=N$) in the super radiant phase \cite{emary03}. From the point of view of the displacement $a^\dag\rightarrow a^\dag+\sqrt{\alpha}$, it follows that, in the thermodynamical limit, the vacuum fluctuations in the field are negligible concerning the interaction with the mirror.

As an alternative way of showing the macroscopic limit of the Dicke system, we present in the appendix A a study of the classical analogue of the present model. This method has been used to study chaos \cite{aguiar92,emary03} and the bifurcation of the equilibrium in the ground state \cite{aguiar91} in the Dicke model. 
It gives reliable results in the large $N$ limit of scalable systems like the Dicke model.

From all of this, it follows that the vibrational motion of the mirror can then work as a probe to the super radiant phase transition. If $\lambda<\lambda_c$, the average occupation $\langle c^\dag c\rangle(t)$ will be time independent according to Hamiltonian (\ref{np}), whereas just above the critical point, the mirror starts to evolve under 
\begin{eqnarray}\label{hm}
H_{\rm{mirror}}=\omega_m c^\dag c+\Omega(c^\dag+c), 
\end{eqnarray}
with $\Omega=-\frac{g_0\lambda^2}{\omega^2}(1-\mu^2)$, and it will cause the average occupation of the vibrational levels to change in time. For example, if the Dicke system is initially in the ground state of $H_{\rm{srp}}^{\rm{Dicke}}$ and the mirror in the vacuum, we will have 
\begin{eqnarray}\label{cc}
\langle c^\dag c\rangle(t)=\frac{2\Omega^2}{\omega_m^2}\left[1-\cos(\omega_mt)\right],
\end{eqnarray} 
for $\lambda>\lambda_c$. 
Of course, it is not really necessary to cool down the mirror at first place. Any initial state would be good because what matters is that $c^\dag c$ is a constant of motion only in the normal phase according to our findings. Consequently,  only in the super radiant phase of the structured bath we will be able to see some dynamics of this observable.
 
\begin{figure}[ht]
\centering\includegraphics[width=0.8\columnwidth]{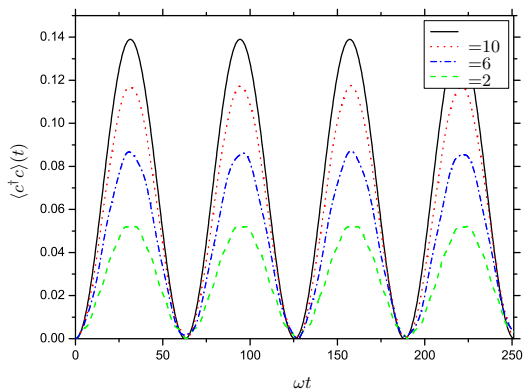}
 \caption{(Color online) Time evolution of mean occupation number $\langle c^\dag c\rangle(t)$ for increasing values of $J$. It is clear the the curves tend to the driven case (\ref{cc}) found in the thermodynamical limit (TL). The system parameters are $\omega=\omega_0=10\omega_m$ and $\lambda=0.6$ (super radiant phase).}
 \label{fig02}
\end{figure}

 The above behavior in the super-radiant phase is numerically evidenced in Fig.\ref{fig02} for the mean occupation number $\langle c^\dag c\rangle(t)$ simulated with Hamiltonian (\ref{hamiltonian}). The system has been initially prepared in the state $|\Psi(0)\rangle=|{\rm{ground}}\rangle_{\rm{Dicke}}\otimes|0\rangle_c$, where $|{\rm{ground}}\rangle_{\rm{Dicke}}$ and $|0\rangle_c$ are the ground states of (\ref{D}) and $H_m=\omega_m c^\dag c$, respectively. Although far from the thermodynamic limit, we can already see that as $J$ increases, the oscillations tend to the regular energy oscillation (\ref{cc}) of the forced harmonic oscillator. The simulation with finite $J$ was performed using standard diagonalization packages. When mapping the atomic Pauli operators to collective angular momentum operators, it is easy to see that the basis for treating the atomic part in the Dicke system has actually dimension $2J+1$, where $J=N/2$. Consequently, cases with $J$ about $15$ are amenable to implementation in current personal computers. There is here, of course, the complication that the basis for the whole system includes also the field and the mirror subsystems. However, since we are simulating just the cases in which the initial state comprises the fundamental state of the Dicke system and the vacuum state of the mirror, the numerical calculations converge quickly without the need to drastically increase the basis dimension.

\begin{figure}[ht]
 \centering\includegraphics[width=0.8\columnwidth]{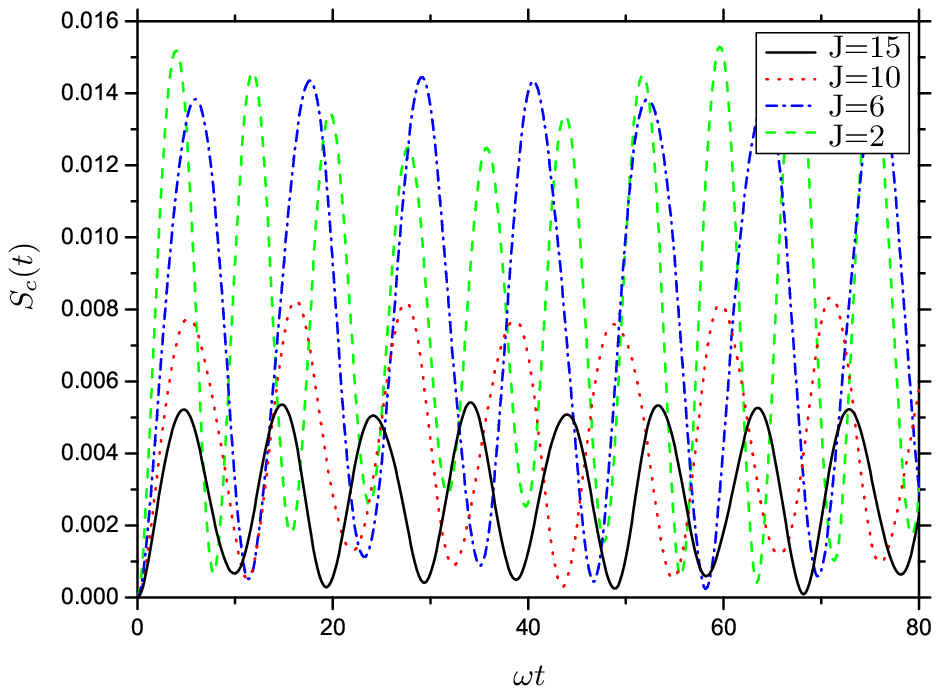}
 \caption{(Color online) Time evolution of the Von Neumann entropy of the mirror state for increasing values of $J$. The system parameters are again $\omega=\omega_0=10\omega_m$ and $\lambda=0.6$ (super radiant phase).}
 \label{fig03}
\end{figure}

According to our findings, the reservoir does not induce decoherence on the mirror in the thermodynamical limit. This can be seem from the fact that there are no products of $a$ or $b$ with $c$ in Hamiltonians (\ref{np}) and (\ref{srp}). Consequently, the initial amount of entropy of the mirror state will not change under the presence of the Dicke system (reservoir). In Fig.\ref{fig03}, we keep the same initial preparation considered in Fig.\ref{fig02}, and we show the numerically evaluated time evolution of the Von Neumann entropy $S_c(t)=-{\rm{Tr}[\rho_c(t)\ln\rho_c(t)}]$ of the mirror state $\rho_c(t)={\rm{Tr}}_{a,J}\rho(t)$, where $\rho(t)$ is the density operator of the whole system. One can clearly see the tendency of the entropy to diminish as $J$ increases, confirming the statement above ($S_c(0)=0$ for initial state of the mirror). 
\section{Conclusions and Final Remarks}\label{final}
We have presented a \emph{new effect} in optomechanics, which is  a super-radiance generated classical driving force on a moving mirror.
We started from the Dicke Hamiltonian and derived an effective Hamiltonian considering harmonic motion of one of the mirrors composing the cavity. The subsequent application of the thermodynamical limit reveals the appearance of a driving force on the moving mirror in the super-radiant phase of the Dicke system, 
 without causing any coherence loss during the time evolution of the mirror state. Although the thermodynamical limit of this system may not be completely under the reach of present technology, we expect that the tendency of the mirror to appear driven in one phase and free in the other may soon be  observed in the regime of few atoms in optomechanical systems of the type reported in \cite{kimble}.

\begin{acknowledgments}
 K.F. wishes to thank M. C. Nemes for helpful discussions. J.P.S. acknowledges financial support  from Coordenação de Aperfeiçoamento de Pessoal de Nível Superior (CAPES). K.F. and F.L.S. acknowledge participation as members of the Brazilian National Institute of Science and Technology of Quantum Information (INCT/IQ). K.F. acknowledges partial support from CNPq  under grant $304041/2007-6$.  F.L.S. also acknowledges partial support from CNPq under grant $303042/2008-7$. 
\end{acknowledgments}
\appendix

\section{A classical analogue of the model}\label{A1}

In this appendix, we present a classical analogue of the model using a time dependent variational method based on coherent states \cite{saraceno}. This method produces for the normalized coherent states representation the following classical Hamiltonian (in units of $\hbar=1=m$): 
\begin{eqnarray}
\mathcal{H} &=& \langle z_a,w,z_c |H|z_a,w,z_c \rangle\nonumber\\
&=&\frac{\omega_0}{2}(q_1^2+p_1^2)-J\omega +  
\frac{\omega}{2}(q_2^2+p_2^2) +  \frac{\omega_m}{2}(q_3^2+p_3^2) \nonumber
\\
&& +\frac{2\lambda}{\sqrt{2J}}\sqrt{2J-\mathcal{H}_1}q_1q_2 - \frac{g_0\sqrt{2}}{4J}
(q_2^2+p_2^2)q_3,
\label{a1}
\end{eqnarray}
where $H$ is given by \eqref{hamiltonian}, $\mathcal{H}_1= (q_1^2+p_1^2)/2$, and the coherent states $|z_a,w,z_c \rangle$
are defined as $|z_a,w,z_c \rangle=|z_a\rangle \otimes |w\rangle \otimes |z_c \rangle$ where $|z_d \rangle = e^{(zd^{\dagger}-\bar{z}d)} |0\rangle (d=a,c)$   and  $|w \rangle= \frac{1}{(1+w\bar{w})^{J}}e^{w J_+}|J,-J\rangle$. The phase space coordinates $(q_1,p_1)$ and  $(q_2,p_2)$  (similarly for $(q_3,p_3)$ with $z_c$) read \footnote{Note that we have exchanged the role of $p_i$ and $q_i$ ($i=1,2$) with respect to reference \cite{aguiar92}.} 
\begin{eqnarray}
q_1&=&\sqrt{\frac{J}{1+w\bar{w}}}(w+\bar{w}). \nonumber\\ 
p_1 &=& \sqrt{\frac{J}{1+w\bar{w}}}\frac{(w-\bar{w})}{i}.\\
q_2&=&\sqrt{\frac{1}{2}}(z_a+\bar{z_a}),\nonumber\\ 
p_2 &=& \sqrt{\frac{1}{2}}\frac{(z_a-\bar{z_a})}{i},\nonumber\\
\end{eqnarray} 

According to (\ref{a1}), the associated classical equations of motion are given by
\begin{eqnarray}
\dot{q}_1 &=& \omega_0 p_1 - \frac{\lambda}{\sqrt{2J}}\frac{p_1q_1q_2}{\sqrt{2J-\mathcal{H}_1}} \nonumber \\
\dot{p}_1  &=&  -\omega_0 q_1 - \frac{2\lambda}{\sqrt{2J}}\sqrt{2J-\mathcal{H}_1}q_2 
+ \frac{\lambda}{\sqrt{2J}}\frac{q_1^2q_2}{\sqrt{2J-\mathcal{H}_1}} \nonumber \\
\dot{q}_2  &=&  \omega p_2  - \frac{g_0\sqrt{2}}{2J} p_2 q_3 \nonumber \\
\dot{p}_2  &=&  -\omega q_2 - \frac{2\lambda}{\sqrt{2J}}{\sqrt{2J-\mathcal{H}_1}}
q_1 +\frac{g_0\sqrt{2}}{2J} q_2q_3  \nonumber \\
\dot{q}_3  &=&  \omega_m p_3 \nonumber \\
\dot{p}_3  &=&  -\omega_m q_3 +  \frac{g_0\sqrt{2}}{4J} (q_2^2+p_2^2).
\label{eqmotion}
\end{eqnarray}

In the macroscopic limit $(N=2J\rightarrow \infty)$, and below critical coupling $\lambda<\lambda_c=\frac{\sqrt{\omega_0\omega}}{2}$ where no bifurcation of the minimum energy fixed point occurs, one can see from the last two equations that the oscillator with frequency $\omega_m$ decouples from the rest of the system and follows free evolution. 

On the other hand, at $\lambda=\lambda_c$ a pitchfork bifurcation occurs for the Dicke model lowest energy fixed point \cite{aguiar91}, such that the stable fixed points for $\lambda\ge\lambda_c$ are  $(q_{10},p_{10})=(\mp\sqrt{2J(1-\mu)},0)$ and  $(q_{20},p_{20})=(\pm \sqrt{4J\frac{\lambda^2}{\omega^2}(1-\mu^2)},0)$. As a kind of first order perturbation theory, we may use this values into the last two equations in \eqref{eqmotion}, and this leads naturally to a forced harmonic oscillator equation  in the limit $N\rightarrow \infty$:
\begin{eqnarray}
\ddot{q}_3 +  \omega_m^2 q_3 =  \left\{ \begin{array}{l}
                      0 , \,\,\,\,\,\,{\rm{if}} \lambda \le \lambda_c ,\\
\sqrt{2}\omega_m g_0 \frac{\lambda^2}{\omega^2}(1-\mu^2), \,\,\,\,\,\,{\rm{if}} \lambda \ge \lambda_c. 
\end{array}\right.\nonumber\\
\label{eqmotion3}
\end{eqnarray}
Now, it is clear that, at least for small $g_0$-coupling, the driving term has the same form as indicated in Sec. \ref{results}, i.e. proportional to $\Omega=-g_0 \frac{\lambda^2}{\omega^2}(1-\mu^2)$ and consistent with the quantum Hamiltonian \eqref{hm}. 

The Dicke model quantum phase transition considering dissipation in the 
a cavity field has been treated in ref. \cite{car_par}, where it has been
shown that the transition point is shifted to $\lambda_c= \frac{1}{2}\sqrt{\omega \omega_0 (1+\frac{\kappa^2}{\omega^2})}$, where $\kappa$ is the loss rate.
Moreover, the fixed point of semiclassical equations results in $z_{a0}=\frac{1}{\sqrt{2}}[q_{20}+ip_{20}]=\pm \sqrt{2J}\frac{\lambda}{\omega-i\kappa}\sqrt{1-\mu^2}$. For small losses ($\kappa \ll \omega$), the right hand side of above Eq. \eqref{eqmotion3} can be substituted by  $\sqrt{2}\omega_m g_0 \frac{\lambda^2}{\omega^2}(1-\mu^2)(1- \frac{\kappa^2}{\omega^2})$. Since the mean photon number in the super radiant ground state is written as $z_a z_a^*=\frac{q_{20}^2+p_{20}^2}{2}$ and this is diminished for $\kappa > 0$, the force acting on the mirror in the super radiant phase is also diminished due to dissipation.  

\end{document}